\documentclass[11pt]{article}
\usepackage{moriond,epsfig}
\usepackage{amsfonts}
\usepackage{amsmath}
\usepackage{amssymb}
\usepackage{amstext}

\def\Journal#1#2#3#4{{#1} {\bf #2}, #3 (#4)}

\def\apj{\em ApJ}
\def\aa{\em A\&A}
\def\araa{\em ARA\&A}

\def\be{\begin{equation}}
\def\ee{\end{equation}}
\def\ba{\begin{array}}
\def\ea{\end{array}}
\def\bea{\begin{eqnarray}}
\def\eea{\end{eqnarray}}
\def\vec#1{\ensuremath{\mathchoice{\mbox{\boldmath$\displaystyle#1$}}
{\mbox{\boldmath$\textstyle#1$}}
{\mbox{\boldmath$\scriptstyle#1$}}
{\mbox{\boldmath$\scriptscriptstyle#1$}}}}

\begin{document}
\vspace*{4cm}

\title{CONSTRAINTS ON \boldmath$(\Omega_\mathrm{M0},\Omega_\Lambda)$\unboldmath
\ FROM STRONG LENSING IN AC\,114}

\author{ G. GOLSE, J.-P. KNEIB \& G. SOUCAIL }

\address{Laboratoire d'Astrophysique de l'Observatoire Midi-Pyr\'en\'ees\\
14, av. E.-Belin, 31400 Toulouse, France}

\maketitle\abstracts{
We use a strong lensing inversion in the cluster of galaxies AC\,114 to
derive constraints on the cosmological parameters $\Omega_\mathrm{M0}$ and
$\Omega_\Lambda$. If it
     is possible to measure spectroscopically the redshifts of {\it
       many} multiple images then one can in principle constrain
     ($\Omega_\mathrm{M0},\Omega_\Lambda$) through ratios of angular diameter
     distances, {\em independently} of any external assumptions. 
	Numerical tests on simulated data show rather good constraints from
     this test. We also
     use an analytic ``pseudo-elliptical'' NFW profile in the simulations,
     following the general new formalism we present. An
     application to AC\,114 favors a flat Universe, an EdS model being
     marginally ruled out.
}

\section{Introduction}

Several independent results seem to converge to a specific cosmological model,
namely an accelerating flat Universe with $\Omega_\mathrm{M0}\simeq 0.3$ and
$\Omega_\Lambda\simeq0.7$. However, there still remain many sources 
of uncertainties with all methods. Thus, any other
independant test to constrain the large scale geometry of the Universe is
important to investigate.

Gravitational lensing has been
considered as a very promising tool for such determinations. Two major
methods are currently used: the statistics of gravitational lenses 
and the cosmic shear variance. 
We focus here on a measurement technique of 
$(\Omega_\mathrm{M0},\Omega_\Lambda)$ using gravitational lensing as a purely
geometrical test of the curvature of the Universe. 

In the case of several sets of multiple 
images, it is possible in principle to constrain the geometry of the Universe,
as suggested by Blandford \& Narayan~\cite{Blandford} and analysed by Link \& 
Pierce~\cite{Link}. Following their method, 
we try to quantify in Sect.~2 what can be
reasonably obtained on $(\Omega_\mathrm{M0},\Omega_\Lambda)$ from accurate
modeling of cluster-lenses, and we apply the test to AC\,114.

Strong lensing effect is very sensitive to the precise projected 
gravitational potential. Yet, for many widespread
profiles, an analytic expression cannot be derived in the elliptical case.
In Sect.~3, we present a general pseudo-elliptical formalism that 
makes possible analytical expressions for lensing quantities. 
This formalism is then applied to the NFW profile.

\section{Cosmological Parameters from Strong Lensing}

	\subsection{Influence of $\Omega_\mathrm{M0}$ and $\Omega_\Lambda$ on
image formation}

In the lens equation $\vec{\theta_\mathrm{S}}=\vec{\theta_\mathrm{I}}
-D_\mathrm{LS}/D_\mathrm{OS}\,\vec{\nabla}\phi(\vec{\theta_\mathrm{I}})$, 
the dependance on the
cosmological parameters is solely contained in the term $E=D_\mathrm{LS}/
D_\mathrm{OS}$. But with a single system of images,
we can only constrain the combination $\sigma_0^2E$, where $\sigma_0$ is the
central velocity dispersion of the lens.

If a gravitational lens shows two systems of multiple images, at redshifts
$z_\mathrm{S1}$ and $z_\mathrm{S2}$, then the ratio $E(z_\mathrm{S2})/
E(z_\mathrm{S1})$ does not depend on $\sigma_0$, so that lower order terms like
$\Omega_\mathrm{M0}$ and $\Omega_\Lambda$ can be probed independently
of the mass normalization. Several observational numbers 
have also to be gathered to derive interesting constraints.
We must get spectroscopic redshifts for the lens and the
sources with a good accuracy, so that $\delta z\simeq0.001$. 
The positions of the different
images have to be obtained very accurately, e.g. with HST images, so that
$\delta\theta\simeq0.01''$. Finally, a strong lensing inversion requires
a precise modeling of the potential model, i.e. considering its type, the
substructures and individual galaxies, as well as the ellipticities of the
different clumps.

Under these conditions, and in a typical configuration ($z_\mathrm{L}=0.3$,
$z_\mathrm{S1}=0.6$ and $z_\mathrm{S2}=5$), we can derive the expected error
bars on the cosmological parameters in two cases (Golse 
{\it et al.}~\cite{Golse}):
\bea
\Lambda\mathrm{CDM}: & \qquad & \delta \Omega_\mathrm{M0}=0.3\pm0.11  
\qquad \delta \Omega_\Lambda=0.7\pm0.23 \\
\mathrm{EdS}: & \qquad & \delta \Omega_\mathrm{M0}=1\pm0.17 
\qquad \quad \delta \Omega_\Lambda=0\pm0.48
\eea

However, these typical values may depend on the choice of the lens 
parameters and on the potential chosen to describe the lens, a problem that we 
investigate below. 

	\subsection{Numerical simulations}

\begin{figure}[!h]
\psfig{figure=IMAGE_NFW.ps,height=2.5in}
\psfig{figure=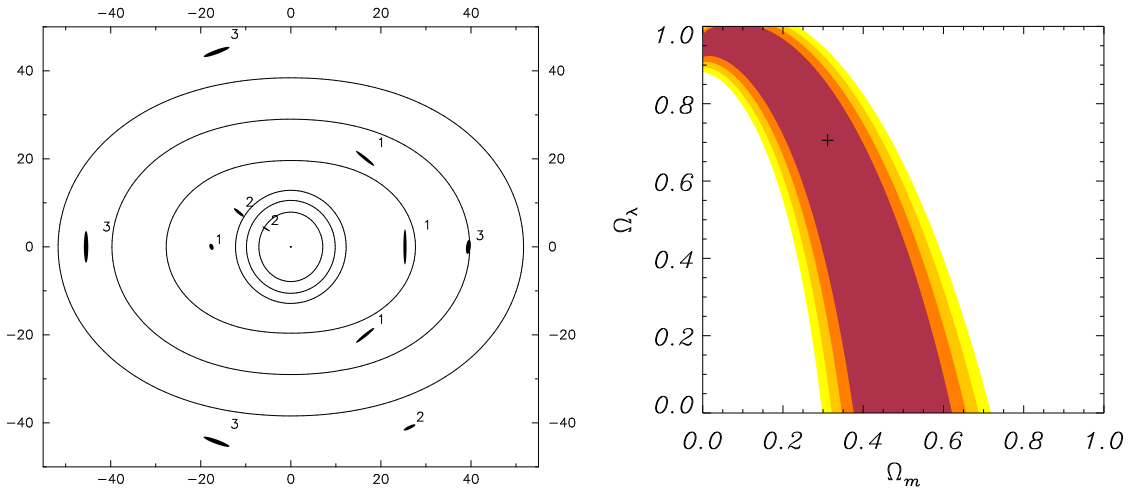,height=2.5in}
\caption{{\bf Left:} Multiple images generated by a pseudo-elliptical 
($\epsilon=0.1$) NFW
cluster at $z_\mathrm{L}=0.3$. Close to their respective critical lines, 3
systems of images are identified: $z_\mathrm{S1}=0.6$, $z_\mathrm{S2}=1$,
$z_\mathrm{S3}=4$. Units are given in in arcseconds. {\bf Right:} $\chi^2$ 
confidence
levels (from 1-$\sigma$ to 4-$\sigma$) obtained from the optimization of the
former lens configuration. The cross (+) represents the original values
$(\Omega_\mathrm{M0}^0,\Omega_\Lambda^0)=(0.3,0.7)$.
\label{Simulation}}
\end{figure}

To create a simulated lens configuration we need to fix some aritrary
values of the cosmological parameters
$(\Omega_\mathrm{M0}^0,\Omega_\Lambda^0)$. The initial data are several sets 
of multiple images at
different redshifts, produced by the numerical code {\it LENSTOOL}
(Kneib~\cite{Kneib93}). With
these observables, we can recover some parameters of the potential
while we scan a grid in the $(\Omega_\mathrm{M0},\Omega_\Lambda)$ plane. The
likelihood of the result is obtained via a $\chi^2$-minimization, 
where $\chi^2$ typically compares
the difference in the images positions to the resolution of the field image.

An ellipticity in the gravitational potential is included in the model, using
analytic lensing expressions introduced
for pseudo-elliptical profiles (Sect.~3), 
and apply this formalism to the NFW mass profile. 

With $(\Omega_\mathrm{M0}^0,\Omega_\Lambda^0)=(0.3,0.7)$,
we generated 3 systems of multiple images (see Fig.~\ref{Simulation}). 
During the optimization process, we
kept fixed the geometrical parameters and recovered the physical ones:
the caracteristic density $\rho_\mathrm{c}$ and the scale radius 
$\theta_\mathrm{s}$. The corresponding
confidence levels on $(\Omega_\mathrm{M0},\Omega_\Lambda)$ are plotted in
Fig.~\ref{Simulation}. This degeneracy is typical of our test, as shown in the
many situations explored in Golse {\it et al.}~\cite{Golse}. 

	\subsection{Application to the clusters of galaxies AC\,114}

\begin{figure}[!h]
\psfig{figure=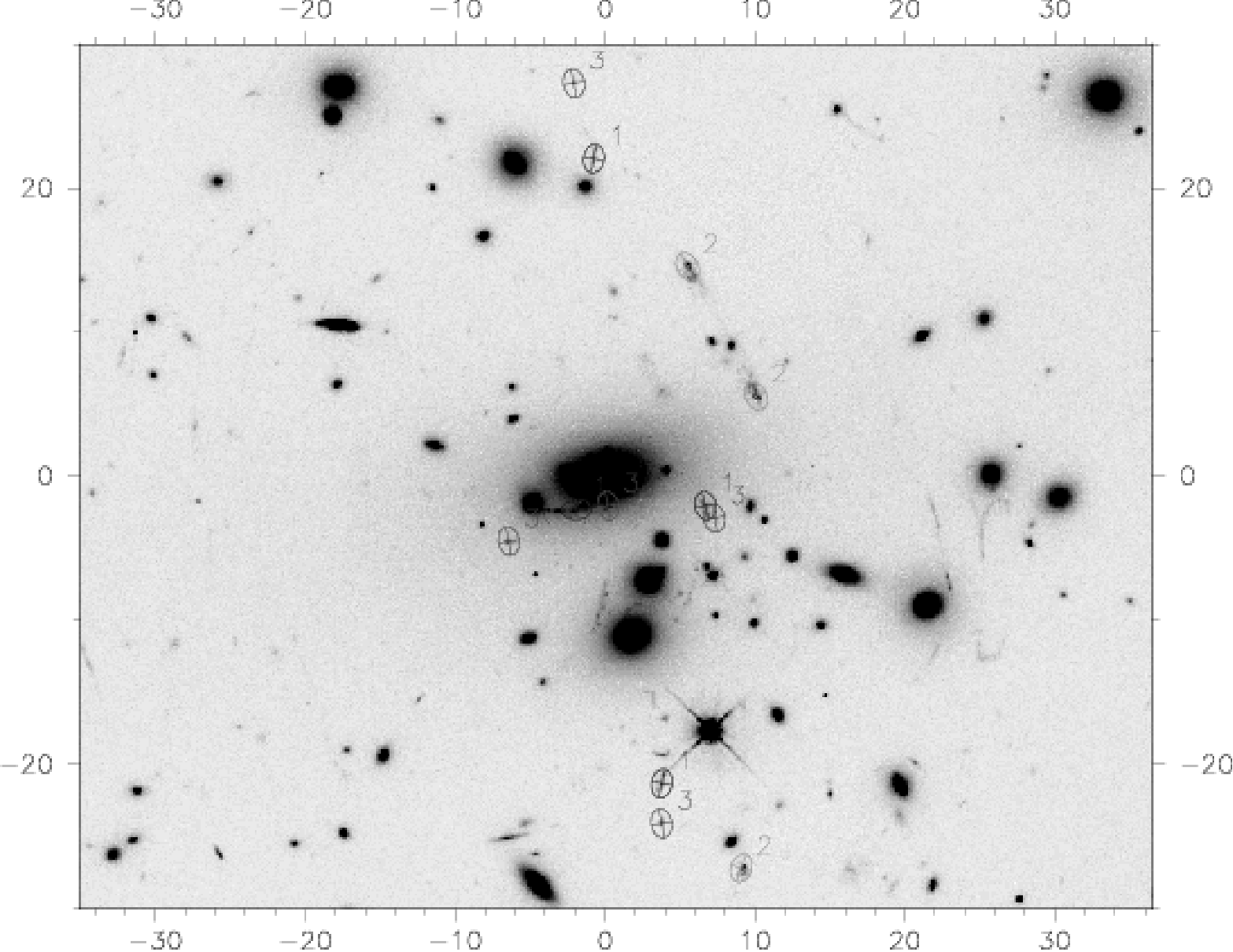,height=2.5in}
\psfig{figure=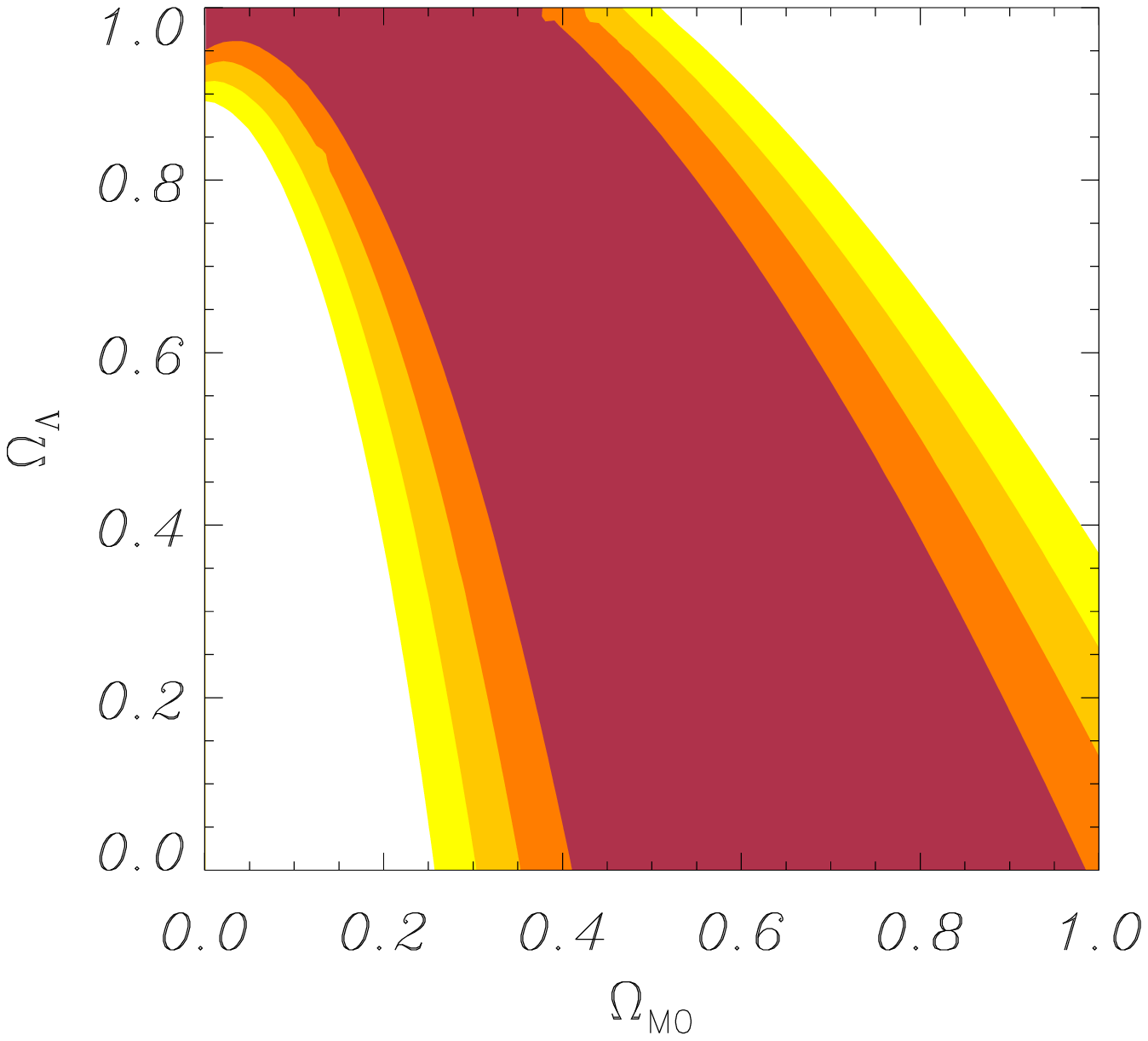,height=2.5in}
\caption{{\bf Left:} {\it HST} image of the center of the cluster AC\,114.
3 systems of multiple images are determined. {\bf Right:} $\chi^2$ confidence
levels (from 1-$\sigma$ to 4-$\sigma$) obtained from the optimization process 
in AC\,114. 
\label{AC114}}
\end{figure}

AC\,114 ($z_\mathrm{L}=0.312$) is particularly well-suited to our test since 
it shows 3 systems of multiple images
with spectroscopically determined redshifts~: $z_\mathrm{S1}=1.691$,
$z_\mathrm{S2}=1.867$ and $z_\mathrm{S3}=3.347$ (Campusano~\cite{Campusano}),
see Fig.~\ref{AC114}.

We find a good fit of all these images positions by considering three
main clumps, and also the individual galaxies. We use elliptical models 
from Hjorth \& 
Kneib~\cite{Hjorth} for all the components of the potential (clumps and 
galaxies). For the optimization, we fix all the parameters
except the core radius $\theta_\mathrm{s}$ and the velocity dispersion 
$\sigma_0$ of the central clump. Moreover, the velocity dispersions of the 
galaxies and other clumps are scaled with respect to $\sigma_0$.

This lensing optimization of the mentionned parameters on a 
$(\Omega_\mathrm{M0},\Omega_\Lambda)$ grid leads to the Fig.~\ref{AC114}
confidence levels on the cosmological parameters. These rather good constraints
favor a flat Universe, including the new standard one 
($\Omega_\mathrm{M0}=0.3$, $\Omega_\Lambda=0.7$), as well as open models with
high matter densities. Note that an EdS model is 
only excluded at the 1-$\sigma$ level.

\section{Pseudo Elliptical Lensing Mass Model: application to the NFW profile}

Cosmological $N$-body simulations of cluster formation indicate the existence
of a universal density profile for dark matter halos (Navarro 
{\it et al.}~\cite{Navarro}). On the other hand, gravitational lensing is an 
ideal tool to constrain the radial structure of collapsed halos like
galaxies and clusters. Mu\~noz {\it et al.}~\cite{Munoz}
introduced a general set of ellipsoidal models. However, as there are no
general analytic expressions for cusped ellipsoidal models, they calculated
the lensing quantities numerically. We propose a new method to introduce
ellipticity in lensing models in a fully analytical way.

We introduce an ellipticity 
$\epsilon$ in the circular lens potential $\varphi(\theta)$. Moreover,
we assume that the radial profile can be scaled by a scale 
radius $\theta_s$, thus making possible to define $x$ as $x=\theta/\theta_s$.
We introduce the ellipticity 
in the expression of the lens potential by substituting $x$ by
$x_\varepsilon=\sqrt{x_{1\epsilon}^2 +
x_{2\epsilon}^2}$, using the following elliptical coordinate system:
\begin{equation}
\left\lbrace
\ba{lcl}
x_{1\epsilon} & = & \sqrt{1-\epsilon} \, x_1 \\
x_{2\epsilon} & = & \sqrt{1+\epsilon} \, x_2 \\
\phi_\epsilon & = & \arctan \left(x_{2\epsilon} / x_{1\epsilon}\right)
\ea
\right.
\label{defin_ell}
\end{equation}

Our method can be used if the potential $\varphi$ (and/or the deflection
angle $\alpha$) and the projected mass
density $\Sigma$ both have analytical expressions in the circular case. 
We can derive easily the corresponding convergence 
$\kappa_\epsilon(\vec{x}) = \displaystyle{\kappa(\vec{x}_\epsilon)+
\epsilon\cos{2\phi_\epsilon}\,\gamma(\vec{x}_\epsilon)}$ 
(see Golse \& Kneib~\cite{GK} for more details).
Similarly, the shear can be written as:
$\gamma_\epsilon^2(\vec{x})=\gamma^2(\vec{x}_\epsilon) + 2\epsilon
\cos{2\phi_\epsilon}\gamma(\vec{x}_\epsilon)\kappa(\vec{x}_\epsilon) 
+ \epsilon^2(\kappa^2(\vec{x}_\epsilon)-\cos^2{2\phi_\epsilon}
\gamma^2(\vec{x}_\epsilon))$.
Finally, the projected mass density is
$\Sigma_\epsilon(\vec{x})=\Sigma(\vec{x}_\epsilon)+\epsilon\cos{2\phi_\epsilon}
(\overline{\Sigma}(\vec{x}_\epsilon)-\Sigma(\vec{x}_\epsilon))$.

We apply this formalism to the NFW profile, for which both the lens potential
(Meneghetti {\it et al}~\cite{Meneghetti}) and the projected mass profile 
(Bartelmann~\cite{Bartelmann}) are known analytically. 

An illustration of some lensed images using is shown in Fig.~\ref{Simulation}. 
In particular it is possible
to form a 5-image configuration. One can estimate precisely (see Golse \& 
Kneib~\cite{GK}) the range of ellipticities for which this model is a good
description of elliptical mass distributions. We shall consider only
$\epsilon\lesssim0.25$, which translates to a limit of 
$\epsilon_\Sigma\lesssim0.4$ for the projected mass density.

\section{Conclusion}

Following the work of Link \& Pierce~\cite{Link}, we discussed a method to
obtain information on the cosmological parameters $\Omega_\mathrm{M0}$ and 
$\Omega_\Lambda$ while reconstructing the lens gravitational potential of 
clusters with multiple image systems at different redshifts.

This technique gives degenerate constraints, with a better precision on the
matter density.
The cluster AC\,114, displaying 3 systems of multiple images, is well-suited
for this method. The optimization process favors a flat Universe, or open
ones with a high matter density. 

Actually the degeneracy depends only on the different redshifts involved that
we will have various sets of when applying the method to real configurations. 
This should lead to a more reduced area of allowed cosmological parameters,
when combining data from different clusters. We
plan to apply this technique to clusters like MS2137-23, MS0440+02, A370, A383 
and A1689.

Strong lensing inversion requires a precise gravitational potential model.
For this reason we propose a new and simple formalism that allows
analytical expressions for the lensing quantities in elliptical models.
We applied this formalism to the NFW profile and estimated the range
of ellipticity ($\epsilon\lesssim0.25$, or  $\epsilon_\Sigma\lesssim0.4$) for
which this model is a good description of elliptical mass distributions.
This will be particularly useful to determine the slope of the central radial
mass profile in clusters of galaxies.

\end{document}